\documentclass[10pt,twocolumn]{article}

\usepackage[utf8]{inputenc}
\usepackage{amsmath}
\usepackage{amssymb}
\usepackage{amsthm}
\usepackage{titlesec}
\usepackage{cite}
\usepackage{color}
\usepackage{units}
\usepackage{booktabs}
\usepackage{graphicx}
\usepackage[colorlinks=false]{hyperref}
\usepackage[format=plain,labelfont=it]{caption}
\usepackage[left=1.5cm,right=1.5cm,top=2cm,bottom=2cm]{geometry}
\usepackage{threeparttable}
\usepackage{multirow} 
\usepackage{adjustbox}
\usepackage{tikz}
\usetikzlibrary{arrows}
\usetikzlibrary{shapes}
\usetikzlibrary{decorations.pathmorphing}
\usetikzlibrary{decorations.pathreplacing}
\usetikzlibrary{decorations.shapes}
\usetikzlibrary{decorations.text}
\usetikzlibrary{positioning, shapes}
\usetikzlibrary{calc}
\tikzset{
block/.style = {draw, fill=white, rectangle, minimum height=3em, minimum width=4.5em},
tmp/.style  = {coordinate}, 
sum/.style= {draw, fill=white, circle, node distance=2cm},
input/.style = {coordinate},
output/.style= {coordinate}
pinstyle/.style = {pin edge={to-,thick,black}}
}
\usepackage[labelformat=empty]{subfig}

\usepackage{tikzit}

\tikzstyle{medium box}=[fill=white, draw=black, shape=rectangle, minimum width=3em, minimum height=3em]
\tikzstyle{junction}=[fill=black, draw=none, shape=circle, minimum size=3pt, inner sep=0pt, outer sep=0pt]

\tikzstyle{arrow}=[->, thick, >=latex]
\tikzstyle{thick dotted line}=[-, loosely dotted, thick]

\titleformat{\section}{\centering\normalfont\scshape}{\Roman{section}.}{5pt}{}
\titleformat{\subsection}{\normalfont\it}{\Alph{subsection}.}{5pt}{}
\titleformat{\subsubsection}{\normalfont\it}{\hspace{4mm}\arabic{subsubsection})}{5pt}{}

\newcommand\infoFootnote[1]{%
  \begingroup
  \renewcommand\thefootnote{}\footnote{#1}%
  \addtocounter{footnote}{-1}%
  \endgroup}

\theoremstyle{remark}
\newtheorem{thm}{Theorem}
\newtheorem{cor}[thm]{Corollary}

\newtheorem{assum}{Assumption}
\newtheorem{prop}[thm]{Proposition}

\newcommand{\R}{\mathbb{R}} 
\newcommand{\N}{\mathbb{N}} 

\newcommand{\Z}{\mathbb{Z}}

\newcommand{\Xc}{\mathcal{X}}

\newcommand{\T}{\mathcal{T}}
\newcommand{\Fc}{\mathcal{F}}

\newcommand{\Uc}{\mathcal{U}}

\newcommand{\Tc}{\mathcal{T}}

\newcommand{\oneb}{\boldsymbol{1}}

\newcommand{\ub}{\boldsymbol{u}}

\newcommand{\xib}{\boldsymbol{\xi}}

\newcommand{\gammab}{\boldsymbol{\gamma}}
\newcommand{\betab}{\boldsymbol{\beta}}

\newcommand{\xb}{\boldsymbol{x}}

\newcommand{\bb}{\boldsymbol{b}}
\newcommand{\cb}{\boldsymbol{c}}

\newcommand{\gb}{\boldsymbol{g}}

\newcommand{\vb}{\boldsymbol{v}}
\newcommand{\wb}{\boldsymbol{w}}

\newcommand{\Ab}{\boldsymbol{A}}
\newcommand{\Bb}{\boldsymbol{B}}

\newcommand{\Lb}{\boldsymbol{L}}

\newcommand{\Xib}{\boldsymbol{\Xi}}
\newcommand{\Ib}{\boldsymbol{I}}

\newcommand{\Pb}{\boldsymbol{P}}
\newcommand{\Qb}{\boldsymbol{Q}}
\newcommand{\Rb}{\boldsymbol{R}}

\newcommand{\Kb}{\boldsymbol{K}}

\newcommand{\Vb}{\boldsymbol{V}}
\newcommand{\Yb}{\boldsymbol{Y}}

\newcommand{\Abc}{\boldsymbol{\mathcal{A}}}
\newcommand{\Bbc}{\boldsymbol{\mathcal{B}}}
\newcommand{\Cbc}{\boldsymbol{\mathcal{C}}}
\newcommand{\Dbc}{\boldsymbol{\mathcal{D}}}
\newcommand{\Ebc}{\boldsymbol{\mathcal{E}}}
\newcommand{\Kbc}{\boldsymbol{\mathcal{K}}}
\newcommand{\Lbc}{\boldsymbol{\mathcal{L}}}

\newcommand{\norm}[1]{\left\lVert#1\right\rVert}

\newcommand{\Enc}{\mathtt{Enc}}

\newcommand{\modu}{\,\,\,\mathrm{mod}\,\,}
\newcommand{\modq}{\,\,\,\mathrm{mod}\,\,q}

\newcommand{\blind}[1]{\textcolor{white}{#1}}

\title{\vspace{-2mm}\bf Secure learning-based MPC via garbled circuit}
\author{Katrine Tjell$^\ast$, Nils Schlüter$^\ast$, Philipp Binfet, Moritz Schulze Darup\vspace{2mm}}
\date{}

\begin{document}

\maketitle

\textbf{\textit{Abstract}.} {\bf 
Encrypted control seeks confidential controller evaluation in cloud-based or networked systems. Many existing approaches build on homomorphic encryption (HE) that allow simple mathematical operations to be carried out on encrypted data. Unfortunately, HE is computationally demanding and many control laws (in particular non-polynomial ones) cannot be efficiently implemented with this technology.

We show in this paper that secure two-party computation using garbled circuits provides a powerful alternative to HE for encrypted control. More precisely, we present a novel scheme that allows to efficiently implement (non-polynomial) max-out neural networks with one hidden layer in a secure fashion. These networks are of special interest for control since they allow, in principle, to exactly describe piecewise affine control laws resulting from, e.g., linear model predictive control (MPC). However, exact fits require high-dimensional preactivations of the neurons. Fortunately, we illustrate that even low-dimensional learning-based approximations are sufficiently accurate for linear MPC. In addition, these approximations can be securely evaluated using garbled circuit in less than 100~ms for our numerical example. Hence, our approach opens new opportunities for applying encrypted control.}
\infoFootnote{$^\ast$K. Tjell and N. Schl\"uter equally share the first authorship. K. Tjell is with the Department of Electronic Systems, Aalborg University, Denmark. E-mail: kst@es.aau.dk. N. Schl\"uter, P. Binfet, and M. Schulze Darup are with the \href{https://rcs.mb.tu-dortmund.de/}{Control and~Cyber-physical Systems Group}, Faculty of Mechanical Engineering, TU Dortmund University, Germany. E-mails:  \href{mailto:nils.schlueter@tu-dortmund.de}{\{nils.schlueter, philipp.binfet, moritz.schulzedarup\}@tu-dortmund.de}. \vspace{0.5mm}}
\infoFootnote{\hspace{-1.5mm}$^\dagger$This paper is a \textbf{preprint} of a contribution to the 60th IEEE Conference on Decision and Control 2021.}

\vspace{-2mm}
\section{Introduction}
Cloud computing and distributed computing are omnipresent in modern control systems such as smart grids, robot swarms, or intelligent transportation systems. While networked control offers exciting features, it also involves privacy and security concerns. Cybersecurity thus becomes an important part of the controller design~\cite{Knowles2015_CIP}. Encrypted control addresses this challenge by providing modified controllers that are able to run on encrypted data~\cite{darup2020encrypted}.
Encrypted controllers have already been realized for various control schemes. For instance, encrypted implementations of (static or dynamic) linear feedback have been proposed in~\cite{Kogiso2015,Farokhi2017,Cheon2018need}. Encrypted distributed controllers can, e.g., be found in~\cite{Farokhi2017,SchulzeDarup2019_LCSS}. Finally, encrypted MPC schemes have been presented in, e.g.,~\cite{SchulzeDarup2018_LCSS,Alexandru2018_CDC,Schlueter2020_CDC}.

For practical applications, secure MPC is of particular interest since networked control systems are often subject to input or state constraints and since optimization-based control can benefit from cloud-based implementations. Hence, encrypted MPC is also in the focus of this paper. More specifically, we propose a substantial extension to the scheme in~\cite{Schlueter2020_CDC} that enhances security, performance, and applicability. 

In order to specify our contributions, we briefly summarize the underlying scheme next and point out its weaknesses. The approach in~\cite{Schlueter2020_CDC} builds on the observation that linear MPC can be exactly represented by tailored max-out neural networks. These networks then build the basis for a (partially) secure two-party implementation. The main weaknesses of the approach are twofold: First, the exact representation of the MPC control law requires high-dimensional preactivations that significantly limit performance and applicability. Second, the scheme requires decryption of some data to evaluate the max-out neurons, which results in a security flaw.

We overcome the weaknesses in~\cite{Schlueter2020_CDC} with two central improvements. Instead of a computationally demanding exact MPC representations, we consider learning-based approximations that allow us to significantly reduce the size of the considered neural networks. Furthermore, we use garbled circuits~\cite{Bellare2012} in combination with secret sharing~\cite{Cramer2015secure}  and show that the resulting architecture enables secure and efficient evaluations of max-out networks.

The remaining paper is organized as follows. Section~\ref{sec:background} provides background on MPC, secret sharing, and garbled circuits. Our main result, i.e., a novel secure implementation of max-out neural networks, is presented in Section~\ref{sec:main}. Finally, Section~\ref{sec:numerical} illustrates our method with a numerical example and Section~\ref{sec:outlook} states conclusions and an outlook. 

\textit{Notation}. We denote the sets of real, integer, and natural numbers (including $0$) by $\R$, $\Z$, and $\N$, respectively. We further denote positive integers by $\N_+$ and define $\N_q$ as $\{0,\dots,q-1\}$ for some $q\in\N_+$. Moreover, we frequently use the modulo operation $z \modq:=z-q\lfloor z/q \rfloor$ and we state congruence modulo~$q$ as $z_1 \equiv z_2 \modq$. In this context, $\lfloor\cdot \rfloor$, $\lceil\cdot \rceil$, and $\lfloor\cdot \rceil$ refer to the floor function, the ceiling function, and rounding to the nearest integer, respectively. Next, for a vector $\vb \in \R^q$, $\max\{\vb\}$ stands for $\max\{\vb_1,\dots,\vb_q\}$. Finally, let $n\in \N_+$ and consider the matrices $\Kb \in \R^{q\times n}$ and $\Pb \in \R^{q \times q}$. Then, 
 $\norm{\Kb}_{\max}:=\max_{i,j }\left|\Kb_{ij}\right|$ and $\|\vb \|_{\Pb}^2:=\vb^\top \Pb \vb.$

\section{Preliminaries and background}
\label{sec:background}

\subsection{Model predictive control via max-out networks}

\textbf{MPC}. Classical MPC builds on solving an optimal control problem (OCP) of the form 
\vspace{-4mm}
\begin{align}
\label{eq:OCP}
V(\xb) := \!\!\!\!\!\min_{\substack{\hat{\xb}(0),...,\hat{\xb}(N)\\ \hat{\ub}(0),...,\hat{\ub}(N-1)}} 
\!\!\!\!\!\!\!\!\!\!\!\!\!\!\!\!\!\! & 
\,\,\,\,\,\,\,\,\,\,\,\,\,\,\,
\| \hat{\xb}(N)\|_{\Pb}^2 \! + \!\! \sum_{\kappa=0}^{N-1} \!\|\hat{\xb}(\kappa)  \|_{\Qb}^2 \!+ \!\| \hat{\ub}(\kappa) \|_{\Rb}^2\!   \\
\nonumber
\text{s.t.} \quad \quad  \hat{\xb}(0)&=\xb, \\
\nonumber
 \hat{\xb}(\kappa+1)&=\Ab\,\hat{\xb}(\kappa) + \Bb \hat{\ub}(\kappa), \qquad\forall \kappa \in \N_N, \\
 \nonumber
\left(\hat{\xb}(\kappa),\hat{\ub}(\kappa)\right) & \in \Xc \times \Uc, \qquad\hspace{16.6mm}\forall \kappa \in \N_N, \\
 \nonumber
 \hat{\xb}(N) & \in \Tc\\[-6.5mm]
 \nonumber
\end{align}
in every time step $k\in\N$ for the current state ${\xb=\xb(k)}$. Here, $N\in\N_+$ refers to the prediction horizon and $\Pb$, $\Qb$, and $\Rb$ are positive (semi\nobreakdash-) definite weighting matrices. The dynamics of the linear prediction model are described by $\Ab\in \R^{n \times n}$ and ${\Bb \in \R^{n\times m}}$. State and input constraints can be incorporated via the polyhedral sets $\Xc$ and $\Uc$. Finally, the terminal set $\T$ allows enforcing closed-loop stability (see \cite{Mayne2000} for details). The resulting control law $\gb:\Fc \rightarrow \Uc$ is defined~as
\begin{equation}
\label{eq:gMPC}
\gb(\xb):=\hat{\ub}^\ast(0),
\end{equation}
where $\Fc$ denotes the feasible set of~\eqref{eq:OCP} and where $\hat{\ub}^\ast(0)$ refers to the first element of the optimal input sequence. We briefly note that one could replace~\eqref{eq:OCP} by the OCP related to robust tube-based MPC (as done in~\cite{SchulzeDarup2018_LCSS,Schlueter2020_CDC}) in order to account for  approximation errors (specified later in Prop.~\ref{Prop1}) and to ensure robust constraint satisfaction. This step is omitted here for ease of presentation.

\textbf{Max-out networks}. In this paper, we exploit learning-based approximations of $\gb(\xb)$ in order to realize secure cloud-based MPC. More specifically, we approximate $\gb(\xb)$ based on max-out neural networks~\cite{Goodfellow2013}. 
For ease of presentation, we will focus on scalar inputs, i.e., $m=1$, throughout the paper, although our approach is not limited to this special case. Hence, we only consider scalar-valued $\gb(\xb)$ that we denote by $g(\xb)$. Moreover, we restrict ourselves to max-out networks of the~form
\begin{equation}
\label{eq:maxout}
\hat{g}(\xb):=\max\{ \Kb \xb + \bb \} - \max \{\Lb \xb + \cb\}
\end{equation}
with $\Kb,\Lb \in \R^{p \times n}$ and $\bb,\cb \in \R^p$, i.e., to a single hidden layer with two neurons that each pool $p \in \N_+$ affine preactivations.

The restriction to networks of the form~\eqref{eq:maxout} is due to three observations. First, according to \cite[Thm.~4.3]{Goodfellow2013}, \eqref{eq:maxout} allows to approximate any continuous function $\tilde{g}:\R^n \rightarrow \R$ arbitrarily well and \eqref{eq:gMPC} is well-known to be continuous. Second, by exploiting the piecewise affine structure of linear MPC \cite{Bemporad2002}, \eqref{eq:maxout} even allows to exactly describe \eqref{eq:gMPC} as specified in \cite{Schlueter2020_IFAC}. Third, as recently noted in \cite{Schlueter2020_CDC}, the form~\eqref{eq:maxout} is beneficial for encrypted implementations.
 
\subsection{Secret sharing and secure two-party computation}
\label{subsec:backSecret}

As mentioned in the introduction, we will use secret sharing \cite{Cramer2015secure} in combination with garbled circuits \cite{Bellare2012} to overcome the security flaws in \cite{Schlueter2020_CDC} and to derive a fully encrypted implementation of~\eqref{eq:maxout}. In the following, we briefly summarize these cryptographic tools.

\textbf{Secret Sharing}. The central idea of secret sharing is quite simple. Assume we intend to perform cloud-based computations on a secret number $z\in \Z$. Then, we can simply divide $z$ into two (or more) random shares and perform the upcoming computations via two (or more) non-colluding clouds. Here, we focus on the special case of two clouds, i.e., two-party computation, that has also been considered in \cite{Schlueter2020_CDC}. Slightly more formally, the essentials of secret sharing and two-party computation can be summarized as follows. Two shares of $z$ are constructed by randomly choosing $z^{(1)}$ from $\N_q$ and subsequently selecting  $z^{(2)} \in \N_q$ such that
\begin{equation}
    \label{eq:zSharesRelation}
    z \equiv z^{(1)} + z^{(2)} \modq,
\end{equation}
where $q \in \N_+$ defines the size of the message space. It can easily be verified that neither $z^{(1)}$ nor $z^{(2)}$ leak information about $z$, and that $z$ can only be reconstructed by combining both shares (as detailed below).

\textbf{Two-party computations}. Remarkably, the shares allow carrying out computations. To see this, we introduce the shorthand notation $[z]:=[z^{(1)},z^{(2)}]$ for shared values. Now, consider two secret numbers $z_1,z_2 \in \Z$ and the corresponding shares $[z_1]$ and $[z_2]$. Then, the number $z_3$ associated with 
\begin{equation}
    \label{eq:sharedAdditions}
[z_3]=[z_1]+[z_2]:=\big[z_1^{(1)}+z_2^{(1)},z_1^{(2)}+z_2^{(2)}\big] \modq
\end{equation}
satisfies $z_3 \equiv z_1+z_2 \modq$. In other words, the shares allow carrying out secure additions. Analogously, secure multiplications and additions with public 
constants $a,b\in\Z$ can be performed. In fact,
\begin{equation}
    \label{eq:sharedOpsWithConsts}
[z_4]=a [z_1]+b:=\big[a z_1^{(1)}+b,a z_1^{(2)} \big] \modq
\end{equation}
is such that $z_4 \equiv a z_1+b \modq$. Finally, even secure multiplications of two secret numbers can be realized. However, this requires the utilization of so-called Beaver triples~\cite{beaver}. A Beaver triple consists of shares $[\alpha]$, $[\beta]$, and $[\gamma]$ from $\N_q$ that satisfy $\alpha \beta \equiv \gamma \modq$. Moreover, the shares are generated by randomly choosing $[\alpha]$ and $[\beta]$ without revealing both shares of $\alpha$, $\beta$, or $\gamma$ to any of the computing parties. Without giving details, we briefly note that various protocols for two-party generation of Beaver triples exist~\cite{Deevashwer2019, Frederiksen2015}. Now, given a Beaver triple, secure multiplications of the form $z_1 z_2$ can be carried out as follows. First, the shares $[\delta]:=[z_1]-[\alpha]$ and $[\epsilon]:=[z_2]-[\beta]$ are computed, where subtractions are defined analogously to additions. Since $[\alpha]$ and $[\beta]$ have been randomly chosen, $[\delta]$ and $[\epsilon]$ contain no information on $[z_1]$ or $[z_2]$. Hence, the two parties can exchange their shares and reveal $\delta$ and~$\epsilon$.~Afterwards,~the~shares
\begin{equation}
    \label{eq:sharedMultiplications}
[z_5]:= [\gamma] + \delta [\beta] + \epsilon [\alpha] + \delta \epsilon 
\end{equation}
can be computed in a distributed and secure fashion. One can easily verify that these shares satisfy the desired multiplicative relation $z_5 \equiv z_1 z_2 \modq$. 

\textbf{Reconstruction}. Obviously, being able to perform secure computations on shared values is useful for encrypted control. However, we eventually need to reconstruct the actual control input and not only a congruent value modulo $q$. To this end, we exploit that the function $\mu:\Z \rightarrow \Z$
\begin{equation}
\label{eq:mu}
\mu (z) := \left\{ \begin{array}{ll}
z-q & \text{if} \,\, z \geq q/2, \\
z & \text{otherwise}
\end{array}\right.
\end{equation}
is a partial inverse of the modulo operation in the sense that the relation $z=\mu( z \modq)$ holds for every 
$$
z \in \Z_q := \left\{ -\lfloor q/2 \rfloor, \dots,  \lceil q/2 \rceil -1 \right\}.
$$

\subsection{Secure circuit evaluations using garbled circuits} 
\label{subsec:garbling}

As specified in Section~\ref{subsec:sharingAffinePreactivations} further below, relations analogous to~\eqref{eq:sharedAdditions}--\eqref{eq:sharedMultiplications} allow to securely evaluate the arguments of the max-operations in~\eqref{eq:maxout}.
However, we do not have a procedure to securely evaluate the max-operations yet. 

\begin{table}[tp]
\caption{Garbled AND-gate with labelled data.} 
\vspace{-3mm}
\label{tab:and}
\begin{center}
\begin{adjustbox}{width=0.4\textwidth}
\begin{tabular}{cccc}
\toprule
\multicolumn{2}{l}{labelled inputs} & outputs & encrypted outputs\\ 
\midrule
\quad$\ell_0^v$ & $\ell_0^w$ & $\ell_0^y$ & $\Enc_{\{\ell_0^v, \ell_0^w\}}(\ell_0^y)$ \\
\quad$\ell_0^v$ & $\ell_1^w$ & $\ell_0^y$ & $\Enc_{\{\ell_0^v, \ell_1^w\}}(\ell_0^y)$\\
\quad$\ell_1^v$ & $\ell_0^w$ & $\ell_0^y$ & $\Enc_{\{\ell_1^v, \ell_0^w\}}(\ell_0^y)$ \\
\quad$\ell_1^v$ & $\ell_1^w$ & $\ell_1^y$ & $\Enc_{\{\ell_1^v, \ell_1^w\}}(\ell_1^y)$ \\
\bottomrule
\end{tabular}
\end{adjustbox}
\end{center}
\vspace{-4mm}
\end{table}

To overcome this issue, we consider so-called garbled circuits \cite{Bellare2012}, which enable secure two-party evaluation of arbitrary Boolean circuits. In the following, we give a brief but intuitive introduction to garbling by considering a Boolean circuit consisting of one AND-gate only. This circuit takes the input bits $v$ and $w$ and returns the output bit $y =\text{AND}(v,w)$. Now, assume $v$ belongs to the first cloud and $w$ to the second one. Then, a garbled circuit allows computing $y$ without revealing any information on $v$ or $w$ to the other cloud (apart from what can be inferred from~$y$). The corresponding procedure can be summarized as follows.

\textbf{Garbling}. The first cloud is the so-called \textit{Garbler}. The Garbler randomly picks two labels (that are associated with $0$ and $1$, respectively) for every input and output in the circuit. Thus, for our simplistic example, the Garbler picks the labels $\ell_0^v, \ell_1^v, \ell_0^w, \ell_1^w, \ell_0^y$, and $\ell_1^y$, where the indices are purely for presentation. The Garbler then creates a truth table for the AND-gate based on the generated labels (see Tab.~\ref{tab:and}). Next, the Garbler encrypts the output column using the corresponding input labels as secret keys leading to, e.g., the  ciphertext $\Enc_{\{\ell_0^v,\ell_0^w\}}(\ell_0^y)$. The garbled circuit is finally generated by randomly ordering the rows of the encrypted truth table
(or by using sophisticated ordering techniques such as point-and-permute \cite{Malkhi2004}). 

\textbf{Communication}. The Garbler sends the garbled circuit to the second party, who acts as the \textit{Evaluator}. Clearly, in order to evaluate the circuit, the Evaluator also needs the garbled inputs in terms of the corresponding labels. Since the labels do not reveal any information on the actual inputs, the Garbler simply sends its labelled input along with the circuit. For instance, in our example, the Garbler sends $\ell_0^x$ if its input is $0$. Since only the Garbler holds the labels and since the Evaluator intends to keep its input secret, it is slightly more complicated to provide the label corresponding to the Evaluator's input. Fortunately, there exist efficient cryptographic protocols that allow to solve this issue. In fact, the Evaluator can infer $\ell_{w}^w$ without revealing $w$ to the Garbler by using a so-called oblivious transfer (OT, \cite{Rabin2005}).

\textbf{Evaluation}. 
Based on the labels $\ell_v^v$ and $\ell_w^w$ and the garbled circuit, the Evaluator can decrypt the related output label (and none of the others). For instance, under the assumption that $v=0$ and $w=1$, the Evaluator obtains $\ell_0^v$ and $\ell_1^w$ and is hence able to decrypt~$\ell_0^y$. Importantly, the Evaluator does not yet know $y$, since the label is a random number. 

\textbf{Revealing}. After evaluating the garbled circuit, the Evaluator holds the resulting output label $\ell_y^y$ but, so far, only the Garbler would be able to interpret it. Depending on the application, these pieces of information can be (re)combined in three different ways. In fact, the plaintext $y$ can either be revealed to the Evaluator, the Garbler, or a third party. We later exploit the first case and therefore comment on this one only. In this scenario, the Garbler simply sends the output labels with the corresponding plaintexts to the Evaluator (which can be done together with the garbled circuit) and the Evaluator uses this information to obtain $y$.

\section{Secure evaluation of max-out networks}
\label{sec:main}

Our main contribution is a secure and efficient two-party implementation of max-out networks as in~\eqref{eq:maxout} based on a tailored combination of secret sharing and garbled circuits. Before presenting our implementation, we specify the goals of the approach and outline the main concept.

\subsection{Goals and concept}

We initially assume that a suitable controller~\eqref{eq:maxout} has been identified by the system operator in terms of $\Kb, \bb, \Lb,$ and $\cb$. Our goal then is to realize a cloud-based evaluation of $\hat{g}$ without revealing the system's state $\xb(k)$, the control actions $\hat{g}(\xb(k))$, or the controller parameters $\Kb, \bb, \Lb$, and $\cb$ to any of the two clouds. To this end, we will use both secret sharing and garbling and try to combine their strengths while avoiding their weaknesses. Namely, as apparent from Section~\ref{subsec:backSecret}, secret sharing is efficient in securely evaluating additions and multiplications while comparisons (as required for max-operations)  are intractable. In contrast, garbled circuits can efficiently  handle comparisons and additions while being very inefficient for  multiplications. As a consequence, we aim for secret sharing-based  evaluations of the affine operations in~\eqref{eq:maxout} while we will exploit garbled circuits for the max-operations.

\subsection{Integer-based reformulation of max-out networks}
\label{subsec:integerBased}

In order to apply secret sharing, we need to reformulate~\eqref{eq:maxout} based on integers. To this end, we simply choose some positive scaling factors $s_1,s_2\in \R_{+}$, define $s_3:=s_1 s_2$, and construct the integers $\xib:=\lfloor s_1 \xb \rceil$, $\Kbc:=\lfloor s_2 \Kb \rceil$,  $\betab:= \lfloor s_3 \bb \rceil$,  $\Lbc:=\lfloor s_2 \Lb \rceil$, and $\gammab:= \lfloor s_3 \cb \rceil$.
Then, the integer-based approximation
\begin{equation}
\label{eq:gIntegerBased}
   \max \{ \Kbc \xib + \betab\} -  \max \{ \Lbc \xib + \gammab \}  \approx s_3 \hat{g}(\xb) 
\end{equation}
holds for sufficiently large scaling factors. More precisely, the following proposition provides an upper bound for possible approximation errors. As a preparation, we introduce the vectors
\begin{equation}
\label{eq:vwIntegerPreactivation}
    \vb:=\Kbc \xib + \betab  \qquad \text{and} \qquad\wb:= \Lbc \xib + \gammab
\end{equation}
that express the results of the affine preactivations in terms of integers.

\begin{prop}
\label{Prop1}
Let $\hat{g}(\xb)$, the scaling factors $s_i$, and $\vb$ and $\wb$ be defined as above.
Further, let $\eta \in \R_+$ and assume that 
\begin{equation}
\label{eq:assumedBounds}
\norm{\xb}_\infty \leq \frac{\eta}{2s_1}, \quad  \norm{\Kb}_{\max} \leq \frac{\eta}{2s_2}, \,\,\, \text{and} \,\,\,  \norm{\Lb}_{\max}\leq \frac{\eta}{2s_2}.\!\!
\end{equation}
Then, deviations between $\hat{g}$ and its integer-based reformulation are bounded above by
\begin{equation}
\label{eq:approxerror}
\!\!\!\left|\,\hat{g}(\xb)-\frac{1}{s_3}\!\left(\max\{\vb\}\!-\max\{\wb\}\right)\right|\leq 
\frac{1}{s_3}\!\left(n \eta+\frac{n}{2}+1\right).
\end{equation}
\end{prop}

\begin{proof}
We first note that the derivation of the upper bound in~\eqref{eq:approxerror} is similar to the one in~\cite[Prop.~2]{Schlueter2020_CDC}. Hence, we can shorten this proof to its essentials. For instance, the corresponding proof in \cite{Schlueter2020_CDC} shows that the max-operations in~\eqref{eq:approxerror} are irrelevant for the desired error bound. As a consequence, the left-hand side of~\eqref{eq:approxerror} is upper bounded~by
\begin{align}
\label{eq:smallerthan}
\norm{\Kb \xb +\bb - \frac{1}{s_3} \vb}_\infty + \norm{\Lb \xb + \cb - \frac{1}{s_3} \wb}_\infty.
\end{align}
We will use the same strategy to evaluate upper bounds for the two terms in~\eqref{eq:smallerthan}. Therefore, we only consider the first term in the following and multiply the resulting bound by $2$ to obtain~\eqref{eq:approxerror}. To this end, we define the deviations
\begin{equation*}
\Delta \xb:=\xb-\frac{1}{s_1}\xib, \;\;
\Delta \Kb:=\Kb-\frac{1}{s_2}\Kbc, \;\; \text{and} \;\;
\Delta \bb:=\bb-\frac{1}{s_3}\betab,
\end{equation*}
and note that these are bounded by 
\begin{equation}
\label{eq:bounds}
\!\!\!\norm{\Delta \xb}_\infty \leq \frac{1}{2s_1}, \;
\norm{\Delta \Kb}_{\max}\leq \frac{1}{2 s_2}, \; \text{and} \;
\norm{\Delta \bb}_\infty\leq \frac{1}{2 s_3}\!\!
\end{equation} 
due to the nearest integer rounding. Based on these deviations, we can rewrite the first expression in~\eqref{eq:smallerthan} as
$$
  \norm{\Kb\xb+\bb-\frac{1}{s_3}\vb}_\infty\!\!
  =\norm{\Delta \Kb \xb + \Kb \Delta \xb - \Delta \Kb \Delta\xb + \Delta \bb}_\infty.
$$
Using subadditivity, we can overestimate the right-hand side by upper bounding the individual terms according to 
$$
\norm{\Delta \Kb \xb}_\infty \leq \| \Delta \Kb \|_\infty \|\xb\|_\infty \leq n \| \Delta \Kb \|_{\max} \|\xb\|_\infty \leq \frac{n \eta}{4 s_1 s_2},
$$
where the two latter relations hold due to~\eqref{eq:assumedBounds} and~\eqref{eq:bounds}.
Proceeding analogously with the three remaining terms, summing up the individual bounds, and multiplying the result by $2$ finally leads to the bound in~\eqref{eq:approxerror}. 
\end{proof}
Remarkably, the error bound~\eqref{eq:approxerror} can be made arbitrarily small by increasing $s_1$ and $s_2$ and hence $s_3$.

\subsection{Secret sharing for affine preactivations}
\label{subsec:sharingAffinePreactivations}
Next, we exploit secret sharing and two-party computation to securely evaluate the integer-based preactivations~\eqref{eq:vwIntegerPreactivation}.
To this end, we assume that each of the two clouds holds one share of the matrices $\Kbc$ and $\Lbc$ as well as the vectors $\betab$ and $\gammab$ that have been uploaded before runtime of the controller. As above, we summarize these shares with $[\Kbc]$, $[\Lbc]$, $[\betab]$, and $[\gammab]$. Additionally, we assume that the sensor computes $\xib$ and related shares $[\xib]$ in every time-step based on the current state $\xb$. With these quantities at hand, we intend to securely compute shares for $\vb$ and $\wb$ as in~\eqref{eq:vwIntegerPreactivation}. In the following, our focus is on the computation of suitable shares $[\vb]$, since $[\wb]$ is computed analogously. 

Clearly, computing $\vb$ requires only scalar-valued multiplications and additions. Therefore, a secure computation based on shares can be carried out according to the procedures in Section~\ref{subsec:backSecret}. In the interest of a compact notation, we extend some of these procedures to matrices next. For instance, we assume that each cloud holds a share of a matrix-valued Beaver triplet $[\Abc]$, $[\Bbc]$, and $[\Cbc]$, where $\Abc$ and $\Bbc$ have been randomly picked from $\N_q^{p\times n}$ and where $\Cbc$ is such that
$$
\Abc \circ \Bbc  \equiv \Cbc \modq
$$
with  ``$\circ$'' denoting the Hadamard (or elementwise) product. Then, the shares $[\vb]$ can be computed as follows.

\begin{prop}
\label{prop:vProperShares}
Let the shares $[\Kbc]$, $[\betab]$, and $[\xib]$ and the Beaver triplet $[\Abc]$, $[\Bbc]$, and $[\Cbc]$ be defined as above and let
$$
[\Xib]:=\begin{pmatrix} [\xib^\top] \\ \vdots \\ [\xib^\top] \end{pmatrix}, \;\;\; [\Dbc]:=[\Kbc]-[\Abc], \;\;\; \text{and} \;\;\; [\Ebc]:=[\Xib]-[\Bbc].
$$
Then, shares for $\vb$  as in~\eqref{eq:vwIntegerPreactivation} can be computed via
\begin{equation}
 \label{eq:properShares}
[\vb]:= \left( [\Cbc] + \Dbc \circ [\Bbc] + \Ebc \circ [\Abc]  + \Dbc \circ \Ebc \right) \oneb + [\betab].
\end{equation}
\end{prop}
 
\begin{proof}
We first note that the Hadamard product allows separating multiplications and additions in the computation of $\vb$. In fact, we obviously have $\vb \equiv (\Kbc \circ \Xib) \oneb + \betab \modq$, where $\Xib$ refers to the recombination of $[\Xib]$. Now, a shared computation of $\Kbc \circ \Xib$ can be carried out analogously to~\eqref{eq:sharedMultiplications} using Beaver's procedure. Clearly, this leads to the expression in the round brackets in~\eqref{eq:properShares}. The remaining operations in~\eqref{eq:properShares} exploit the procedures in \eqref{eq:sharedAdditions} and \eqref{eq:sharedOpsWithConsts}.
\end{proof}

\subsection{Boolean circuit for max-out}
\label{subsec:boolean_maxout}

After evaluating the affine preactivations using secret sharing, the first cloud holds $\vb^{(1)}$ and $\wb^{(1)}$ and the second cloud holds $\vb^{(2)}$ and $\wb^{(2)}$. All shares are, by construction, contained in~$\N_q$ and, hence, non-negative. However, the vectors  $\vb$ and $\wb$ may also contain negative numbers. Thus, before evaluating the max-operations, we need to reconstruct $\vb$ and~$\wb$. Because $[\vb]$ (and $[\wb]$) represent proper shares of $\vb$ (and $\wb$) according to Proposition~\ref{prop:vProperShares}, we can easily derive the following statement.
\begin{cor}
\label{cor:reconstruction}
Assume $\Kbc \xib + \betab \in \Z_q^p$. Then,
\begin{equation}
\label{eq:vReconstruction}
  \vb_i = \mu \big( \vb_i^{(1)} + \vb_i^{(2)} \modu q \big)  
\end{equation} 
for every $i \in \{1,\dots,p\}$.
\end{cor}
Clearly, the reconstructed vectors $\vb$ and $\wb$ should not be revealed to any of the two clouds. Therefore, the reconstruction~\eqref{eq:vReconstruction} has to be included in the Boolean circuit, which is later garbled. More precisely, we will design two garbled circuits for $\nu:=\max\{\vb\}+r_1 \modq$ and $\omega:=\max\{\wb\}+r_2 \modq$, where the role of the random numbers  $r_1, r_2 \in \N_q$ will be clarified further below in Section~\ref{subsec:architecture}. Since the functionality of both circuits is equivalent, we only discuss the design of the $\nu$-circuit. Obviously, a Boolean circuit requires Boolean input data, and we thus need bit-wise representations of $\vb^{(1)}$, $\vb^{(2)}$, and $r_1$ here. In order to provide this data, the following assumption is made.

\begin{assum}
\label{assum:ql}
The modulus $q$ is of the form $q=2^l$ for some $l\in\N_+$.
\end{assum}

\textbf{Bit-wise representation}. Due to this assumption, the vectors $\vb^{(1)},\vb^{(2)} \in \N_q^p$ and the scalar $r_1\in \N_q$ can be represented in terms of $l$-bit unsigned numbers. In order to formalize this observation, we note that there exist unique matrices $\Vb^{(1)},\Vb^{(2)} \in \{0,1\}^{p \times l}$ such that the relations
$$
\vb_i^{(1)}= \sum_{j=1}^l 2^{l-j} \Vb_{ij}^{(1)}  \quad \text{and} \quad \vb_i^{(2)}= \sum_{j=1}^l 2^{l-j} \Vb_{ij}^{(2)} 
$$
hold for every $i\in\{1,\dots,p\}$. Since $\Vb^{(1)}$ and $\Vb^{(2)}$ are Boolean and since they uniquely represent $\vb^{(1)}$ and $\vb^{(2)}$,  we use them as inputs for our Boolean circuit. Analogous observations hold for $r_1$ but are not required for the following discussion. Now, the first task of the circuit is to compute $\vb_i^{(1)}+\vb_i^{(2)}$ based on the corresponding entries in $\Vb^{(1)}$ and $\Vb^{(2)}$ and, subsequently, to evaluate~\eqref{eq:vReconstruction}.

\begin{figure}[tp]
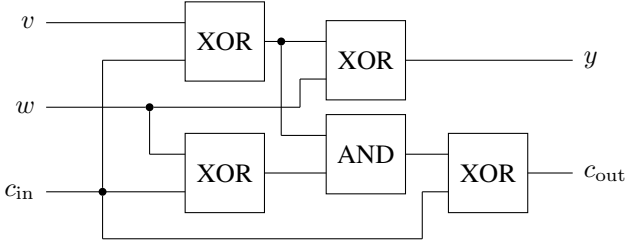

    \centering
\ctikzfig{fa}
    \caption{A Boolean circuit realizing a full adder (FA).}
    \label{fig:FA}
\end{figure}
\begin{figure}[tp]
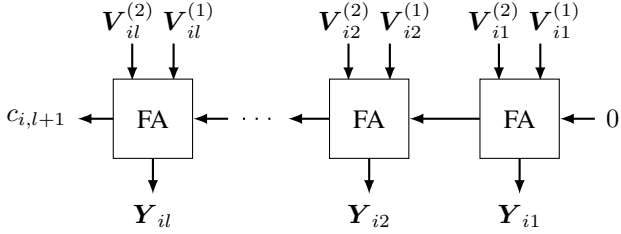

    \centering
\ctikzfig{rca}
    \caption{A ripple-carry adder for adding $l$-bit numbers.}
    \label{fig:RCA}
      \vspace{-4mm}
\end{figure}

\textbf{Bit-wise addition and reconstruction.} In principle, the addition can easily be carried out using standard ripple-carry adders based on $l$ full adders (FA) as illustrated in Figures~\ref{fig:FA} and~\ref{fig:RCA}. However, one has to take into account that properly representing $\vb_i^{(1)}+\vb_i^{(2)}$ might require an additional bit. In fact, this case arises if the last carry bit ($c_{i,l+1}$ in Fig.~\ref{fig:RCA}) evaluates to $1$. At this point, one might be tempted to extend the bit-wise representation of the computed sum by this carry bit. Yet, as apparent from~\eqref{eq:vReconstruction}, the addition is followed by a modulo operation with $q=2^l$. Clearly, whenever we enter this operation with an unsigned number with at least $l$ bits, it will simply return the $l$ least significant bits and cut off all remaining bits. Hence, a possible extension by the mentioned carry bit would immediately be undone by the modulo operation and is thus meaningless. More precisely, let the $i$-th row of the matrix $\Yb \in \{0,1\}^{p \times l}$ reflect the $l$ least significant bits of the sum $\vb_i^{(1)}+\vb_i^{(2)}$. Then,
$$
\sum_{j=1}^l 2^{l-j} \Yb_{ij} = 
\vb_i^{(1)}+\vb_i^{(2)} \modu q. 
$$
Now, in order to reconstruct $\vb_i$ according to~\eqref{eq:vReconstruction}, it remains to apply $\mu$ as in~\eqref{eq:mu}. Interestingly, for $q=2^l$ and $z \in \N_q$, the bit-wise interpretation of~\eqref{eq:mu} simply reflects the conversion from unsigned to signed numbers. Therefore, we obtain
$$
-2^{l-1} \Yb_{i1} + \sum_{j=2}^l 2^{l-j} \Yb_{ij} = \mu \left( \sum_{j=1}^l 2^{l-j} \Yb_{ij} \right)
$$
using the two's complement convention. In other words, evaluating~\eqref{eq:mu} within the circuit does not require any computations but just a reinterpretation of the bit-wise number format. We study two trivial numerical examples in Table~\ref{tab:bitComputations} to illustrate the computations up to this point.

\begin{table}
\caption{Bit-wise evaluation of addition, modulo $q$, and~$\mu$  for two numerical examples with $l=3$ and $c_{i,l+1} \in \{0,1\}$.}
\vspace{-3mm}
\label{tab:bitComputations}
\normalsize
\begin{center}
    \begin{tabular}{ccr}
\toprule
\!\!\!\!var./op.\! & \!\!val.\! & \!\!bit-wise\\
\midrule 
$\blind{+}\vb_1^{(1)}$ & $3$ & $0$  $1$  $1$ \\
$+\vb_1^{(2)}$ & $6$ & $1$  $1$  $0$ \\
\midrule 
$=$ & $9$ & \!($1$) $0$  $0$  $1$ \\
\!\!\!$\mod q$ & $1$ & $0$  $0$  $1$ \\
$\mu(\cdot)$ &  $1$ & $0$  $0$  $1$ \\
\bottomrule
\end{tabular}\qquad \begin{tabular}{ccr}
\toprule
\!\!\!\!var./op.\! & \!\!val.\! & \!\!bit-wise\\
\midrule 
$\blind{+}\vb_2^{(1)}$ & $\blind{+}5$ & $1$  $0$  $1$ \\
$+\vb_2^{(2)}$ & $\blind{+}2$ & $0$  $1$  $0$ \\
\midrule 
$=$ & $\blind{+}7$ & \!($0$) $1$  $1$  $1$ \\
\!\!\!$\mod q$ & $\blind{+}7$ & $1$  $1$  $1$ \\
$\mu(\cdot)$ &  $-1$ & $1$  $1$  $1$ \\
\bottomrule
\end{tabular}
\end{center}
\vspace{-4mm}
\end{table} 

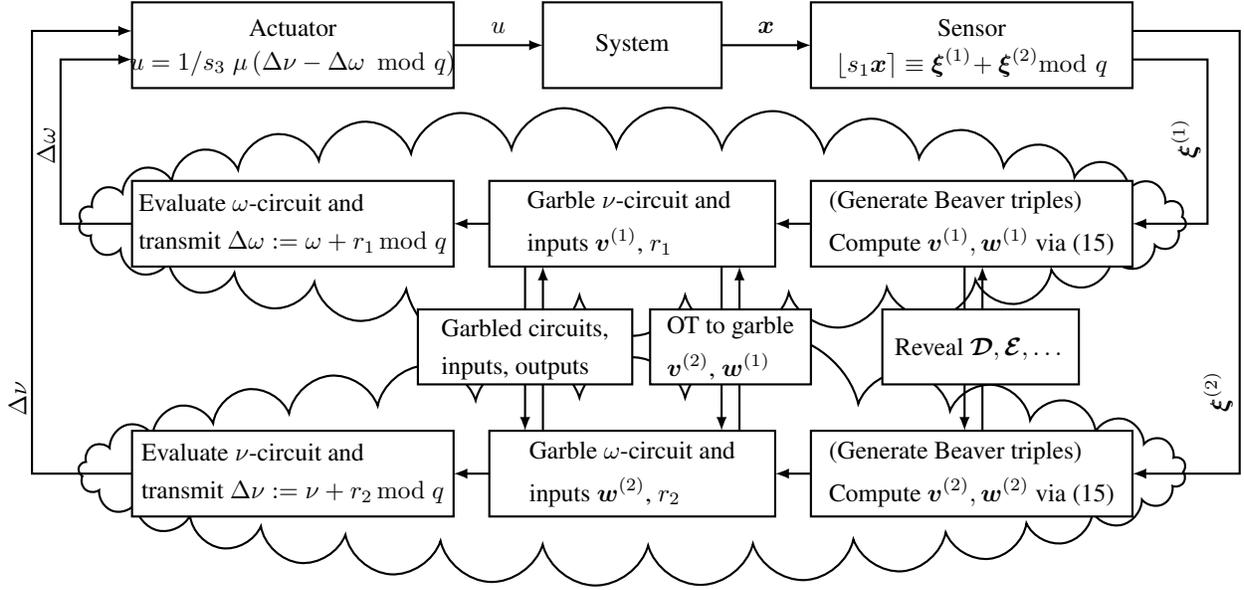
\begin{figure*}[tp]
	\centering
	\begin{tikzpicture}[thick,scale=0.95, every node/.style={scale=0.95}]
	
	\draw [thick] (-3,0.9) rectangle (1.5,2.1);
	\node at (-0.75,1.75) {Actuator};
    \node at (-0.75,1.25) {$u=1/s_3\; \mu \left(\Delta \nu-\Delta \omega \modq \right)$};
	
	\draw [thick, ->, >=latex] (1.5,1.5) to (2.75,1.5);	
	\node at (2.125,1.75) {$u$};
	
	\draw [thick] (2.75,0.9) rectangle (5.25,2.1);
	\node at (4,1.5) {System};	
	
	\draw [thick, ->, >=latex] (5.25,1.5) to (6.5,1.5);	
	\node at (5.875,1.75) {$\xb$};
	
	\draw [thick] (6.5,0.9) rectangle (11,2.1);
	\node at (8.75,1.75) {Sensor};
	\node at (8.75,1.25) {$\lfloor s_1 \xb \rceil\equiv \xib^{(1)}\!+\xib^{(2)}\mathrm{mod}\,\,q$};
	
	\node[cloud, cloud puffs=32.7, cloud ignores aspect, minimum width=15.5cm, minimum height=3.2cm, align=center, draw, rotate=180] (cloud) at (4, -0.97) {};
	
	\draw [thick, ->, >=latex] (11,1.3) -| (12.05,1.3) |- (11,-1);
	\node [rotate=90] at (11.75,0.2){$\xib^{(1)}$};
	
	\draw [thick, fill=white] (6.5,-1.6) rectangle (11,-0.4);
	\node (V) at (8.75,-1){$
	\begin{aligned}
	&\text{(Generate Beaver triples)}\\
	&\text{Compute $\vb^{(1)},\wb^{(1)}$ via~\eqref{eq:properShares}}
	\end{aligned}
	$};	
	
	\draw [thick, ->, >=latex] (6.5,-1) to (6,-1);
	
	\draw [thick,fill=white] (2,-1.6) rectangle (6,-0.4);
	\node at (4,-1) {$
	\begin{aligned}
	    &\text{Garble $\nu$-circuit and } \\
	    &\text{inputs $\vb^{(1)}$, $r_1$}
	\end{aligned}
	$};	
	
	\draw [thick, ->, >=latex] (2,-1) to (1.5,-1);
	
	\draw [thick, fill=white] (-3,-1.6) rectangle (1.5,-0.4);
	\node at (-0.75,-1){$
	\begin{aligned}
	    &\text{Evaluate $\omega$-circuit and } \\
	    &\text{transmit $\Delta \omega:=\omega+r_1\,\mathrm{mod}\;q$}
	\end{aligned}
	$};
    
	\draw [thick, ->, >=latex] (-3,-1) -| (-4,1.3) |- (-3,1.3);
	\node [rotate=90] at (-4.2,0.1) {$\Delta \omega$};
	
	\node[cloud, cloud puffs=32.7, cloud ignores aspect, minimum width=15.5cm, minimum height=3.2cm, align=center, draw] (cloud) at (4, -4.47) {};
	 
	\draw [thick, ->, >=latex] (11,1.7) -| (12.5,1.7) |- (11,-4.5);	
	\node [rotate=90] at (12.2,-3.35){$\xib^{(2)}$};

	\draw [thick, fill=white] (6.5,-5.1) rectangle (11,-3.9);
	\node (V) at (8.75,-4.5){$
    \begin{aligned}
    	&\text{(Generate Beaver triples)}\\
	    & \text{Compute } \vb^{(2)},\wb^{(2)} \text{ via}~\eqref{eq:properShares}
	\end{aligned}
	$};
	
	\draw [thick, ->, >=latex](8.65,-1.6) to (8.65,-3.9);
	\draw [thick, ->, >=latex](8.9,-3.9) to (8.9,-1.6);
	\draw [thick, fill=white] (7.5,-3.25) rectangle (10.25,-2.2);
	\node at (8.875,-2.75){$\text{Reveal $\Dbc,\Ebc,\dots$}
	$};
	
	\draw [thick, ->, >=latex] (6.5,-4.5) to (6,-4.5);
	
	\draw [thick,fill=white] (2,-5.1) rectangle (6,-3.9);
	\node at (4,-4.5) {$
	\begin{aligned}
	    &\text{Garble $\omega$-circuit and} \\
	    &\text{inputs $\wb^{(2)}$, $r_2$}
	\end{aligned}
	$};	
	
	\draw [thick, ->, >=latex] (5.25,-1.6) to (5.25,-3.9);
	\draw [thick, ->, >=latex] (5.5,-3.9) to (5.5,-1.6);
	\draw [thick, fill=white] (4.25,-3.25) rectangle (6.5,-2.2);
	\node at (5.375,-2.75){$
	\begin{aligned}
	    &\text{OT to garble} \\
	    &\text{$\vb^{(2)}$, $\wb^{(1)}$}
	\end{aligned}
	$};
	
	\draw [thick, ->, >=latex] (2.5,-1.6) to (2.5,-3.9);
	\draw [thick, ->, >=latex] (2.75,-3.9) to (2.75,-1.6);
	\draw [thick, fill=white] (1,-3.25) rectangle (4,-2.2);
	\node at (2.5,-2.75){$
	\begin{aligned}
	&\text{Garbled circuits,}\\
	&\text{inputs, outputs}
	\end{aligned}
	$};
	
	\draw [thick, ->, >=latex] (2,-4.5) to (1.5, -4.5);
		
    \draw [thick,fill=white] (-3,-5.1) rectangle (1.5,-3.9);
	\node at (-0.75,-4.5) {$
	\begin{aligned}
	&\text{Evaluate $\nu$-circuit and}\\
	&\text{transmit $\Delta \nu:=\nu+r_2\,\mathrm{mod}\;q$}
	\end{aligned}
	$};	
	
	\draw [thick, ->, >=latex] (-3,-4.5) -| (-4.4,1.7) |- (-3,1.7);
	\node [rotate=90] at (-4.6,-3.45) {$\mathrm{\Delta \nu}$};

	\end{tikzpicture}
	\caption{Overall architecture of the proposed scheme based on secret sharing and garbled circuits.
	}
	\vspace{-4mm}
	\label{fig:encryptedControl2clouds}
\end{figure*}

\textbf{Max-operations}. Having reconstructed all $\vb_i$, it remains to compute $\max \{\vb\}$. This can be done by successively evaluating max-of-two-operations in a tournament-like fashion with $\lceil \log_2(p) \rceil$ rounds (i.e., quarterfinals, semifinals, etc.). We illustrate the bit-wise evaluation of these operations by exemplarily investigating $\max\{\vb_1,\vb_2\}$. Obviously,
\begin{equation}
\label{eq:maxOfTwo}
   \max\{\vb_1,\vb_2\} = \left\{\begin{array}{ll}
\vb_1 & \text{if} \,\,\vb_1 \geq \vb_2, \\
\vb_2 & \text{otherwise}.
\end{array}\right. 
\end{equation}
Now, the two cases in~\eqref{eq:maxOfTwo} can also be identified based on the sign-bit $\sigma$ of $\vb_1- \vb_2$. In fact, the first case refers to $\sigma=0$ and the second to $\sigma=1$. Thus, the maximum can be expressed as $(1-\sigma) \vb_1 + \sigma \vb_2$. Fortunately, this sum can be computed without relying on full-adders. In fact, the $j$-th bit of this sum simply results from
\begin{equation}
\label{eq:xorMulBits}
\text{XOR}\left(\text{AND}\left(\Yb_{1j},\text{NOT}(\sigma) \right),\text{AND}\left(\Yb_{2j},\sigma\right)\right).
\end{equation}
Repeating the previous steps for the remaining entries of $\vb$ and the remaining rounds of the ``tournament'' completes the circuit-based evaluation of $\max\{\vb\}$.

\textbf{Randomization}.
It only remains to add the random number $r_1$ modulo $q$. To this end, we note that
$$
\nu = (\max\{\vb\}  \modq ) + r_1 \modq.
$$
As above, the bit-wise evaluation of $\max\{\vb\}  \modq $ simply reflects a conversion from signed to unsigned numbers. Adding $r_1$ can finally be realized analogously to Figure~\ref{fig:RCA}, where we can (again) ignore the $(l+1)$-th carry bit since the modulo $q$ operation follows.

\subsection{Garbling and reconstruction of control actions}

\textbf{Garbled circuits}. In order to securely evaluate the designed circuit, we use garbling as briefly introduced in Section~\ref{subsec:garbling}. We note, in this context, that the presented circuit only contains AND-, XOR-, and NOT-gates. Now, without giving details, securely evaluating XOR- and NOT-gates does not require any garbling \cite{Kolesnikov2008}. Hence, garbling as in Table~\ref{tab:and} is only required for one AND-gate per FA and two AND-gates per operation~\eqref{eq:xorMulBits}. Zooming out, we require $l$ FAs per addition (and per subtraction). Further, each max-of-two-operation requires one subtraction and $l$ operations of the form~\eqref{eq:xorMulBits}. Finally, by assuming that $p$ is likewise a power of $2$, carrying out the ``tournament'' requires $p-1$  max-of-two-operations. Thus, the circuit contains in total
\begin{equation}
\label{eq:numberOfAndGates}
    p l+3(p-1)l+l=(4p-2) l
\end{equation}
AND-gates, where the three summands on the left-hand side reflect the computations of~\eqref{eq:vReconstruction}, the evaluation of $\max\{\vb\}$, and the addition of $r_1$, respectively. Remarkably, the size of the garbled circuit solely depends on the dimension $p$ of the max-out network and the chosen bit-length $l$.

The clouds use the garbled circuit(s) to securely compute $\nu:=\max\{\vb\}+r_1 \modq$ and $\omega:=\max\{\wb\}+r_2 \modq$, respectively.  More precisely, the first cloud acts as a Garbler for the $\nu$-circuit, which is subsequently evaluated by the second cloud. The roles are inverted for the $\omega$-circuit. As a result, the overall computations are symmetrically allocated between both clouds.

\textbf{Reconstructing control actions}.
After evaluating the garbled circuits, the first cloud holds~$\omega$ and the second one~$\nu$. We show next how to use this data to reconstruct (an approximation of) $\hat{g}(\xb)$ at the actuator. To this end, the first cloud sends $\Delta \omega:=\omega + r_1 \modq $ and the second cloud transmits $\Delta \nu:=\nu + r_2 \modq $. The actuator then computes and applies
\begin{equation}
\label{eq:reconstructControlActions}
  u(k)=\frac{1}{s_3}\mu \left(\Delta \nu- \Delta \omega \modq \right),  
\end{equation}
where we note that both random values $r_i$ cancel out.
Formally, we require $\max\{\vb\}-\max\{\wb\} \in \Z_q$ for a correct evaluation of this final step. However, this condition is typically significantly less restrictive than the one in Corollary~\ref{cor:reconstruction} since the resulting control inputs are restricted to $\Uc$ anyway (apart from approximation errors).

\subsection{Overall architecture and security guarantees}
\label{subsec:architecture}

After having introduced all building blocks of our scheme, we briefly comment on their interplay and resulting security guarantees. In this context, the secure evaluation of the controller can be subdivided in an offline and online phase.

\textbf{Offline}. In the offline phase, the system operator first selects $p$ and then obtains $\Kb, \Lb, \bb$, and $\cb$ by training the max-out network~\eqref{eq:maxout}. Afterwards, $\Kbc, \Lbc, \betab,$ and $\gammab$ are formed by choosing suitable scaling factors $s_1$ and $s_2$. Finally, $q=2^l$ is specified and the shares $\Kbc^{(i)}$, $\Lbc^{(i)}$, $\betab^{(i)}$, and $\gammab^{(i)}$ are generated and transmitted to the $i$-th cloud.

\textbf{Online}. The online procedure is illustrated in Figure~\ref{fig:encryptedControl2clouds}. In every time step, the sensor measures $\xb(k)$ and computes $\xib$. It then generates the shares $\xib^{(i)}$ and sends them to cloud~$i$. Now, the two clouds first exploit~\eqref{eq:properShares} (and the analogue for $[\wb]$) to securely compute shares of $\vb$ and $\wb$. To this end, they generate two sets of matrix-valued Beaver triples $[\Abc]$, $[\Bbc]$, and $[\Cbc]$ using, e.g., the procedure in \cite{Frederiksen2015}. Next, each cloud generates a random number $r_i$, garbles the circuit for one max-out neuron and labels its own inputs to the circuit (i.e., $\vb^{(1)}$ and $r_1$ for the first cloud). After that, it transmits the circuit together with the output table to the other cloud. Simultaneously, the remaining input labels (i.e., for $\vb^{(2)}$ in the $\nu$-circuit) are obtained through OT. Afterwards, each cloud evaluates the received garbled circuit, derives its output, and transmits the result masked by the random $r_i$ to the actuator. Finally, the actuator reconstructs $u(k)$ according to \eqref{eq:reconstructControlActions} and applies it.

\textbf{Security}. To specify security guarantees, we make the (realistic) assumption that the two clouds are non-colluding and honest but curious. In other words, the two clouds will strictly follow the specified protocols without interchanging further data, but they may try to infer information from the given data. Clearly, our security goal is to prevent the latter.
In this context, we first note that the evaluation of the affine preactivations is perfectly secure (i.e, information theoretically secure) by the use of secret sharing \cite{Cramer2015secure}. Next, garbled circuits are proven to be secure against semi-honest adversary \cite{Hazay2010} which fits our assumption about the clouds. Finally, we mask the computed outputs of the max-out neurons by the random numbers $r_i$, which is equivalent to secret sharing and, therefore, enjoys perfect security. In summary, the clouds cannot obtain any information about $\xib$, $\Kbc$, $\Lbc$, $\betab$,  $\gammab$, $\vb$, $\wb$, $\max\{\vb\}$, $\max\{\wb\}$, or the resulting~$u$. We note, however, that generating new Beaver triples and garbled circuits for each evaluation of $u(k)$ is essential for~security.

\section{Numerical benchmark}
\label{sec:numerical}

We illustrate our approach for the standard double integrator example with the system matrices
$$
\Ab=\begin{pmatrix}
1 & 1 \\ 0 & 1
\end{pmatrix} \quad \text{and} \quad \Bb=\begin{pmatrix}
0.5 \\1
\end{pmatrix}
$$
and the constraints $\Xc=\{ \xb \in \R^2 \,|\, |\xb_1| \leq 25, \, |\xb_2| \leq 5 \}$ and $\Uc=\{ u \in \R \, |\, |u| \leq 1 \}$.
The predictive controller associated with~\eqref{eq:OCP} is specified as follows. We set $N=15$, $\Qb=\Ib$, and $\Rb=0.01$ and choose $\Pb$ as the solution of the discrete-time algebraic matrix Riccati equation
$$
\Ab^\top (\Pb- \Pb\,\Bb\,(\Rb+\Bb^\top \Pb\,\Bb )^{-1} \Bb^\top \Pb)\,\Ab - \Pb + \Qb = \boldsymbol{0}.
$$
Finally, $\Tc$ is chosen as the largest subset of $\Xc$ for which the linear quadratic regulator can be applied without violating the state or input constraints.

\begin{figure}[tp]
	\centering
	\includegraphics[trim=4.2cm 9.5cm 3.5cm 10cm, clip=true,width=.95\linewidth]{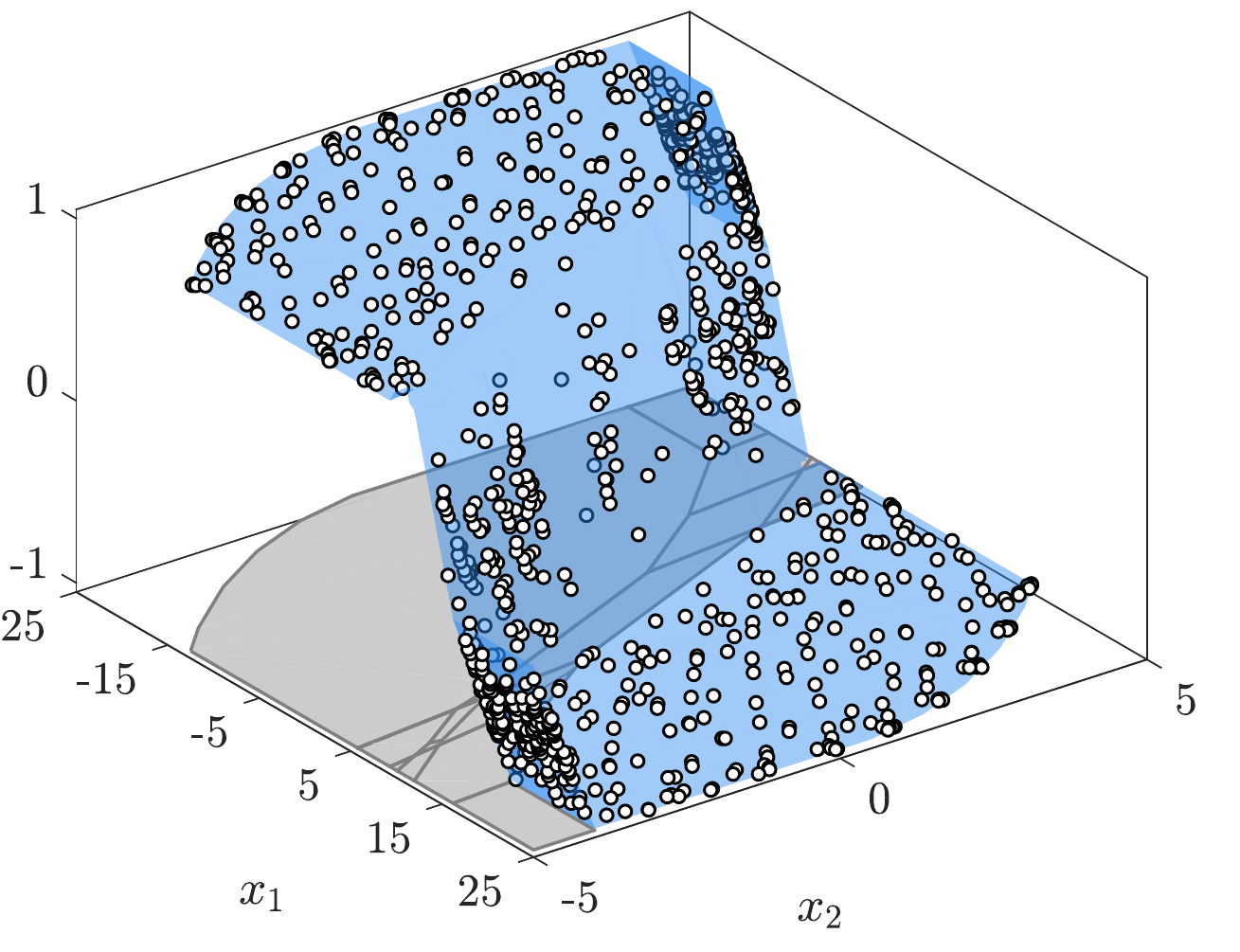}
	\caption{Illustration of the trained max-out neural network (blue) and a few sample points $\big( \xb^{(i)} ,g(\xb^{(i)}) \big)$ (white) over the partition (gray) induced by~\eqref{eq:maxout}. 
	}
	\label{fig:uapprox}
	\vspace{-4mm}
\end{figure}

Next, in order to train the neural network, we sampled $M=6000$ states $\xb^{(i)}$ in the feasible set $\Fc$, evaluated $g(\xb^{(i)})$, and solved the nonlinear program
\begin{align*}
    \min_{\substack{\Kb,\bb,\Lb,\cb}}\;\sum_{i=1}^{M} 
    &\norm{ \max\{\Kb \xb^{(i)}\! +\bb\}-\max\{\Lb\xb^{(i)}\!+\cb\} -g(\xb^{(i)}) }_2^2
\end{align*}
locally, where we fixed the size $p$ to either $8$ or $16$. The resulting mean squared errors (MSE) with respect to~\eqref{eq:gMPC} are listed in Table~\ref{tab:results}. For $p=8$, the local optimizers (rounded to two digits after the decimal point) are given by
\begin{align*} 
\Kb^\top\!\!=
&\begin{pmatrix}   
   -0.07 \!\!\!&\!\!\!
   \blind{-}0.31 \!\!\!&\!\!\! 
    \blind{-}0.01 \!\!\!&\!\!\!
   -0.01 \!\!\!&\!\!\!
    \blind{-}0.31 \!\!\!&\!\!\!
    \blind{-}0.01 \!\!\!&\!\!\!
    \blind{-}0.07 \!\!\!&\!\!\! 
   -0.31
   \\
   -0.52 \!\!\!&\!\!\!
   -0.32 \!\!\!&\!\!\!
   -0.30 \!\!\!&\!\!\!
   -0.40 \!\!\!&\!\!\!
   \blind{-}0.68 \!\!\!&\!\!\!
   \blind{-}0.52 \!\!\!&\!\!\!
   -0.41 \!\!\!&\!\!\!
   -0.64
\end{pmatrix},\\
\Lb^\top\!\!=
&\begin{pmatrix}
    \blind{-}0.31 \!\!\!&\!\!\!   
    \blind{-}0.12 \!\!\!&\!\!\!
   -0.07 \!\!\!&\!\!\!  
   -0.31 \!\!\!&\!\!\!   
   -0.31 \!\!\!&\!\!\!  
   -0.08 \!\!\!&\!\!\!   
    \blind{-}0.01 \!\!\!&\!\!\!  
    \blind{-}0.08   
    \\
    \blind{-}0.68 \!\!\!&\!\!\!
    -0.03 \!\!\!&\!\!\!
    -0.52 \!\!\!&\!\!\!
    \blind{-}0.36 \!\!\!&\!\!\!
    -0.64 \!\!\!&\!\!\!
    \blind{-}0.48 \!\!\!&\!\!\!
    \blind{-}0.52 \!\!\!&\!\!\!
    \blind{-}0.01 
\end{pmatrix}, \\
\bb^\top \!\!=
&\begin{pmatrix}
\hspace{0.35mm}
   \blind{-}0.37 \!\!\!&\!\!\!
   \blind{-}4.60 \!\!\!&\!\!\!
   \blind{-}0.50 \!\!\!&\!\!\!
   \blind{-}0.54 \!\!\!&\!\!\!
   \blind{-}0.60 \!\!\!&\!\!\!
   \blind{-}1.67 \!\!\!&\!\!\!
   -1.14 \!\!\!&\!\!\!
   \blind{-}0.39
\end{pmatrix}, \\
\cb^\top\!\!=
&\begin{pmatrix}
\hspace{0.35mm}
    \blind{-}0.40 \!\!\!&\!\!\!
   -0.88 \!\!\!&\!\!\!
   -1.37 \!\!\!&\!\!\!
   -4.61 \!\!\!&\!\!\!
   -0.61 \!\!\!&\!\!\!
   -1.39 \!\!\!&\!\!\!
   -0.67 \!\!\!&\!\!\!
  \blind{-} 0.40
\end{pmatrix}.
\end{align*}
An illustration of the corresponding function $\hat{g}(\xb)$ is depicted in Figure~\ref{fig:uapprox}.

Now, in order to prepare the secure two-party implementation of the learned controller, we require a suitable integer-based representation of $\hat{g}(\xb)$. In this context, we mainly need to choose the scaling factors $s_1$ and $s_2$ as well as the size of the message space $q$. A suitable choice should preclude overflow. Hence, the assumption in Corollary~\ref{cor:reconstruction} needs to be satisfied for all $\xb \in \Fc$. For practical applications, it is useful to start with the choice of $q=2^l$, since $l$ determines the number format in terms of bits. Here, we consider $l\in\{16,32\}$. Interestingly, fixing $l$ leads to an upper bound for $s_3$, i.e., for the product $s_1 s_2$. To see this, we initially note that the proof of Proposition~\ref{Prop1} implies
$$
\norm{\Kb\xb+\bb-\frac{1}{s_3} ( \Kbc \xib + \betab) }_\infty \leq \frac{1}{2 s_3}\left(n \eta+\frac{n}{2}+1\right):=\Delta. 
$$
Based on this bound, one can show that the implication
$$
\|\Kb \xb + \bb \|_\infty< \frac{q}{2 s_3}-\Delta \quad \implies \quad  \Kbc \xib + \betab \in \Z_q^p
$$
holds. Hence, the assumption in Corollary~\ref{cor:reconstruction} is satisfied if
\begin{equation}
 \label{eq:s3Condition}
   s_3< \frac{2^{l-1}}{\max_{\xb \in \Fc}  \|\Kb \xb + \bb \|_\infty  + \Delta}, 
\end{equation}
where we used $q=2^l$. Now, we can easily evaluate $\max_{\xb \in \Fc} \|\Kb \xb + \bb \|_\infty$ (and $\max_{\xb \in \Fc} \|\Lb \xb + \cb \|_\infty$) based on linear programming. 
However, precisely computing $\Delta$ is difficult, at this point, since it depends on $s_3$ and $\eta$. Fortunately, we will usually be able to realize a small error bound $\Delta$. Therefore, we can simply choose a reasonable overestimation for $\Delta$ and justify (or discard) this choice afterwards. 

Here, we assume $\Delta \leq 1$. The resulting upper bounds $s_{3,\max}$ in dependence of $p$ and $l$ are listed in Table~\ref{tab:results}.
Subsequently, we can freely choose $s_1$ and $s_2$ as long as  $s_3=s_1 s_2$ satisfies the corresponding bound. For instance, we choose $s_1=20$ and $s_2=100$ for $p=8$ and $l=16$, where the ``asymmetric'' choice reflects the fact that $s_2$ applies to a matrix. Having fixed the scaling factors, we choose the smallest $\eta$ such that \eqref{eq:assumedBounds} holds (for every $\xb \in \Fc$). Finally, we can easily verify $\Delta \ll 1$ for all considered combinations of $p$ and $l$. Further, we can quantify the actual quantization errors, i.e., the left-hand side in~\eqref{eq:approxerror}, by computing the MSE based on the sampled states. The results are given in Table~\ref{tab:results}.

Now, regarding the secure controller evaluation according to Figure~\ref{fig:encryptedControl2clouds}, we note that the quantization, the secret sharing, and the garbled circuits are determined by the scaling factors $s_i$, the size of the message space $q$, and the quantities $p$ and~$l$. In particular, the latter quantities determine the size of the garbled circuit according to~\eqref{eq:numberOfAndGates} and the corresponding numbers of AND-gates are listed in Table~\ref{tab:results}. For the actual garbling, we use $128$-bit numbers for each label in the circuit. Further, output labels for each gate are encrypted by the hash of the sum of the associated inputs using SHA-256.

Finally, $1$-out-of-$2$ OT is implemented based on ElGamal encryption. With this setup, a single evaluation of the encrypted controller on an Intel Core i5 with 2.50GHz leads, on average, to the computation times in Table~\ref{tab:results}.
We note, in this context, that the generation of random numbers (including the Beaver triples) and latency has not been taken into account yet.

\setlength{\tabcolsep}{4.5pt}
\begin{table}[t]
\caption{Error estimations and key data for the example.}
\label{tab:results}
\centering
\begin{threeparttable}
\begin{tabular}{ccccccc}
\toprule
$p$ & MSE$^\dagger$ & $l$ & $s_{3,\max}$ & MSE$^\ddagger$ & \#AND & $t_{\mathrm{avg}}$ \\
\midrule
\multirow{2}{*}{\!\!\!$\blind{0}8$}  & \multirow{2}{*}{\!\!\!$18.57\cdot10^{-6}$\!\!\!}         & $16$ & $2.23\cdot10^3$ & $4.37\cdot10^{-5}$\!\! & $\blind{1}480$ & $\blind{1}79$ ms\!\!  \\
                      &                                  & $32$ & $1.51\cdot10^8$ & $5.06\cdot10^{-9}$\!\! & $\blind{1}960$ & $167$ ms\!\! \\
                      \midrule
\multirow{2}{*}{\!\!\!$16$} & \multirow{2}{*}{\!\!\!$\blind{0}1.99\cdot10^{-6}$\!\!\!} & $16$ & $2.21\cdot10^3$ & $6.45\cdot10^{-5}$\!\! & $\blind{1}992$ & $170$ ms\!\!  \\
                      &                                  & $32$ & $1.49\cdot10^8$ & $2.46\cdot10^{-6}$\!\! & $1984$ & $348$ ms\!\!  \\
\bottomrule
\end{tabular}
\begin{tablenotes}
\vspace{.5mm}
    \item[$^{\dagger}$] mean squared error of the max-out approximation w.r.t.~\eqref{eq:gMPC}
    \item[$^{\ddagger}$] mean squared error of the quantization w.r.t.~\eqref{eq:maxout}
    \end{tablenotes}
\end{threeparttable}
\vspace{-4mm}
\end{table}

\section{Conclusions and Outlook}
\label{sec:outlook}
We presented a novel secure implementation of linear MPC over a two-cloud architecture. The key observation motivating our approach is that max-out neural networks are well-suited to approximate piecewise affine control laws while providing a structure that allows for an efficient and secure implementation. In fact, the affine preactivations can be evaluated by additive secret sharing and the subsequent evaluation of the max-operations is realized with garbled circuits. Due to this tailored setup, we avoid inefficient multiplications within the garbled circuits. Improvements compared to existing encrypted MPC schemes lie in reduced evaluation times and a more efficient usage of the involved clouds. More precisely, significantly larger max-out networks~\eqref{eq:maxout} are required in \cite{Schlueter2020_CDC} (because learning is not considered) and~\cite{SchulzeDarup2018_LCSS} evaluates the control law partially at the sensor.

Future research will address two aspects. First, max-out networks of the form~\eqref{eq:maxout} can, in principle, approximate arbitrary (continuous) functions, and they consequently may support secure implementations of (some) nonlinear MPC schemes. Second, we aim for more realistic implementations of our approach including latency, dropouts, and more complex control laws.

\section*{Acknowledgement}
The majority of this work was carried out during an academic visit of Katrine Tjell in the Control and Cyberphysical Systems group at TU Dortmund University. Financial support by the German Research Foundation (DFG) under the grant SCHU 2940/4-1 is gratefully acknowledged.


\end{document}